\documentclass[epjST,nopacs]{svjour}

\usepackage{amsmath}
\usepackage{graphics}
\begin{document}
\title{Hyperspherical Treatment of Strongly-Interacting Few-Fermion Systems in One Dimension}

\author{Artem~G.~Volosniev\inst{1}\fnmsep\thanks{\email{artem@phys.au.dk}} \and Dmitri~V.~Fedorov\inst{1} \and Aksel~S.~Jensen\inst{1} \and Nikolaj~T.~Zinner\inst{1}}
\institute{Department of Physics and Astronomy, Aarhus University, 
DK-8000 Aarhus C, Denmark}
\abstract{
We examine a one-dimensional two-component fermionic system in a trap, assuming that all particles
have the same mass and interact through
a strong repulsive zero-range force. First we show how a simple system of three 
strongly interacting particles in a harmonic trap can be treated using the hyperspherical 
formalism. 
Next we discuss the behavior of the energy 
for the $N$-body system.
} 
\maketitle
\section{Introduction}
\label{intro}
 Recent advances in the preparation of quasi-one-dimensional 
few-fermion samples
in their ground states~\cite{serwane,zurn} showed the need 
for a thorough theoretical description of such systems.   
This need has driven several groups to provide a numerical 
analysis of small systems in a harmonic trap~\cite{gharashi,sowinski,lindgren}.
Unfortunately, a numerical analysis is not reliable 
for large samples with strong interaction where 
the effects beyond the mean field need to be included.
This specific problem was addressed in Ref.~\cite{volosniev} where 
the recipe to calculate the spectrum close to the 
infinite repulsion limit (or hard core limit) was given. 

In the present paper we continue investigation of strongly interacting 
one-dimensional systems. 
First we consider three harmonically 
trapped spinless fermions of two types using 
the hyperspherical formalism~\cite{macek}.
This allows us to show that the wave function of three spinless 
fermions can be used to determine a leading order correction to the energy close 
to the infinite repulsion limit. This conclusion coincides with the result 
of Ref.~\cite{volosniev}. 
Next we offer a new detailed explanation of this result for more particles. 
\section{Three Particles}
\label{sec:1}

We first consider three harmonically trapped spinless fermionic particles two of type 
$A$ (described with coordinates $y_1,y_2$) and one of type $B$
(coordinate $x_1$). The interaction between the particles is 
assumed to be of zero range, which means that due to 
the Pauli principle
particles of type $A$ do not interact 
with each other but only with particle $B$. 
The Hamiltonian for such a system is written as
\begin{equation}
H=-\frac{\hbar^2}{2m}\frac{\partial^2}{\partial x_1^2}
-\frac{\hbar^2}{2m}\frac{\partial^2}{\partial y_1^2}
-\frac{\hbar^2}{2m}\frac{\partial^2}{\partial y_2^2}
+g\delta(x_1-y_1)+g\delta(x_1-y_2) +m\omega^2\frac{x_1^2+y_1^2+y_2^2}{2},
\label{ham}
\end{equation}
where the mass, $m$, is the same for all particles, $\omega$ is the frequency 
of the oscillator trapping potential, and $g$ is the strength 
of the interparticle repulsion. For this problem we shall use
the oscillator units, i.e. all  
lengths are in units of the oscillator 
length~$\sqrt{\hbar/m\omega}$ and
energies are in units of the
trap oscillator energy $\hbar \omega$.
The interaction strength, $g$, becomes
dimensionless in units of $\sqrt{\omega \hbar^3/m}$.
We transform this Hamiltonian from Cartesian 
coordinates to the Jacobi set of coordinates 
with $x=(y_1-y_2)/\sqrt{2}$, $y=\sqrt{\frac{2}{3}}(\frac{y_1+y_2}{2}-x_1)$,
$z=\sqrt{\frac{1}{3}}(y_1+y_2+x_1)$, which yields 
 \begin{equation}
H=-\frac{1}{2}\frac{\partial^2}{\partial z^2}
-\frac{1}{2}\frac{\partial^2}{\partial x^2}
-\frac{1}{2}\frac{\partial^2}{\partial y^2}
+\sqrt{2}g\left[\delta(\sqrt{3}y+x)
+\delta(\sqrt{3}y-x)\right]
+\frac{x^2+y^2+z^2}{2}.
\end{equation}
First we note that the 'center-of-mass' part ($z$) is separable,
so we focus exclusively on the $x,y$ part. Since 
the wave function must be antisymmetric under the $y \to -y$ 
transformation ($A$ particles are identical fermions)
and parity is conserved it is enough
to consider only the $x>0,y>0$ region.
 
To solve the problem we adopt the hyperspherical formalism~\cite{macek,esben}
that has been proven to be useful for three-body 
problems in one dimension~\cite{lindgren,kartavtsev,zinner,harshman}.
In hyperspherical coordinates ($x=r\cos(\phi), y=r\sin(\phi)$) 
the Hamiltonian reads
 \begin{equation}
H=-\frac{1}{2}\frac{\partial^2}{\partial z^2}+\frac{z^2}{2}
-\frac{1}{2r}\frac{\partial}{\partial r}r\frac{\partial}{\partial r}
-\frac{1}{2r^2}\frac{\partial^2}{\partial \phi^2}
+\frac{g}{\sqrt{2}r}\delta(\phi-\pi/6)
+\frac{g}{\sqrt{2}r}\delta(\phi+\pi/6)
+\frac{r^2}{2}.
\end{equation}
To find the eigenstates of this Hamiltonian 
we write the wave function 
in the following form $\Psi_{k}=\frac{\psi_k(z)}{\sqrt{r}}\sum_{i=1} f_i(r)\Phi_i(\phi,r)$,
where $\phi_k(z)$ is the $k$th solution to the harmonic oscillator potential. 
The angular functions,
$\Phi_i(\phi,r)$, are chosen as the normalized solutions of the Schr{\"o}dinger
equation ($0<\phi<\pi/2$, $0<r<\infty$) at fixed $r$
 \begin{equation}
\left(-\frac{1}{2}\frac{\partial^2}{\partial \phi^2}
+\frac{gr}{\sqrt{2}}\delta(\phi-\pi/6)\right)\Phi_i(\phi,r)=E_i(r)\Phi_i(\phi,r),
\label{phi_eq}
\end{equation}
and the radial functions, $f_i(r)$, solve the infinite system of coupled ordinary differential equations
($E$ is now measured from the $k$th energy of the harmonic oscillator):
 \begin{equation}
\left(-\frac{1}{2}\frac{\partial^2}{\partial r^2} +\frac{E_i(r)-1/8}{r^2} +\frac{r^2}{2} -E
\right)f_i(r)=\sum_j\left(Q_{ij}+P_{ij}\frac{\partial}{\partial \rho}\right)f_j(r),
\label{phi_r}
\end{equation}
where $Q_{ij}=\frac{1}{2}\langle \Phi_i|\frac{\partial^2}{\partial r^2}|\Phi_j \rangle_{\phi}$
and $P_{ij}=\langle \Phi_i|\frac{\partial}{\partial r}|\Phi_j \rangle_{\phi}$.
It is worth noting that the system trapped in the harmonic oscillator and 
the corresponding free system have the same set of $\Phi_i$ and correspondingly the 
same couplings $P_{ij}, Q_{ij}$.
The angular equation~(\ref{phi_eq}) has the solutions 
\[ 
\Phi_i = 
 N_i(r)
\left\{ \begin{array}{c c}
     
   -\sin(\mu_i(\phi-\pi/2)) ; \qquad \qquad  \pi/6<\phi<\pi/2 \\
 \left\{
	\begin{array}{l l}
 \frac{\sin(\mu_i\pi/3)}{\cos(\mu_i\pi/6)}\cos(\mu_i\phi);& \quad \text{odd parity} \quad 0<\phi<\pi/6\\
     \frac{\sin(\mu_i\pi/3)}{\sin(\mu_i\pi/6)}\sin(\mu_i\phi); & \quad \text{even parity} \quad 0<\phi<\pi/6
  \end{array} \right.
  \end{array} \right.
\]
where the normalization factor 
$N_i(r)=\sqrt{\frac{3\mu_i}{4\mu_i\pi \mp 2\mu_i\pi\cos(\mu_i\pi/3)\pm3\sin(\mu_i\pi/3)-
3\sin(2\mu_i\pi/3)}}$, the upper sign corresponds to odd parity and 
the lower sign to even parity solutions. 
Also we defined $\mu_i=\sqrt{2E_i}$ where
$E_i$ is chosen to reproduce the discontinuity 
of the derivative of the wave function which arises due to the 
delta function potential at $\phi=\pi/6$. This condition is 
satisfied if $E_i$ 
solves the following equations
for odd parity eigenstate
\begin{equation}
\mu_i \cos(\mu_i\pi/2)+gr\sqrt{2}\cos(\mu_i\pi/6)\sin(\mu_i\pi/3)=0,
\end{equation}
and for even parity eigenstates
\begin{equation}
\mu_i \sin(\mu_i\pi/2)+gr\sqrt{2}\sin(\mu_i\pi/6)\sin(\mu_i\pi/3)=0.
\end{equation}

From now on we focus on the ground state solution 
with strong interaction, i.e. $1/g\ll1$; however
the presented procedure is completely general and can be easily applied 
for excited states. The ground state has odd parity 
and the angular wave function for strong repulsion allows us 
to write a $1/g$ expansion,
$\Phi_i(\phi,R)=a_i(\phi)+\frac{1}{gr}b_i(\phi)+O(1/g^2)$, which yields
$Q_{11}\simeq1/g^2$. Since the solutions with and without $Q_{11}$ give
an upper and lower bounds on the exact energy~\cite{hornos} we can use the lowest 
adiabatic potential alone to describe the energy up to the order $1/g$.
The lowest adiabatic potential is determined by $\mu_1$:
\[ \mu_1(r) = \left\{
	\begin{array}{l l}
3-\frac{27}{\sqrt{2}\pi}\frac{1}{gr} +o(1/gr)& ,r \gg 1/g\\
    1+O(gr) & ,r \ll 1/g
  \end{array} \right.\]
Contribution from small distances is highly suppressed due to the
fermionic nature of the $A$ particles and will not be considered below.
The corresponding equation for the energy takes the form
\begin{equation}
\left(-\frac{1}{2}\frac{\partial^2}{\partial r^2} +\frac{35}{8 r^2}+\frac{r^2}{2}+
V(r) -E\right)f(r)=0,
\label{req}
\end{equation}
where $V(r)=-\frac{81}{\sqrt{2}\pi g r^3}$ (at small distances the potential
becomes regular, but this region can be neglected to linear order in $1/g$). 
We write down the solution to eq.~(\ref{req}) that is regular at zero 
using the Green's function
\begin{equation}
f(r)=R(E,r)+\int_0^r \mathrm{d}r' g_E(r,r') V(r') f(r'),
\label{rfunc}
\end{equation}
where 
\begin{align}
g_E(r,r')=-\frac{\Gamma(3-E/2)\Gamma(-2+E/2)}{\Gamma(2+E/2)}\left[R(E,r)I(E,r')-I(E,r)R(E,r')\right], \\
R(E,r)=e^{-r^2/2} r^{7/2} L_{-2+E/2}^3(r^2), \qquad I(E,r)=e^{-r^2/2} r^{7/2} U(2 -E/2, 4, r^2),
\end{align}
 where $\Gamma(x)$ is the Gamma function,
$U$ is the Tricomi confluent hypergeometric function,
and $L$ is the
associated Laguerre polynomial.
The energy, $E$, is then determined from the condition
of vanishing wave function at infinity
\begin{equation}
1-\frac{\Gamma(3-E/2)\Gamma(-2+E/2)}{\Gamma(2+E/2)}\int_0^\infty \mathrm{d}r' I(E,r)V(r')=0,
\end{equation}
which to linear order in $1/g$ gives 
\begin{equation}
E=4-\frac{27}{\sqrt{2}\pi g}  \int_0^{\infty} \frac{R^2(4,R)}{R^3}\mathrm{d}R =4-\frac{81}{8 \sqrt{2\pi} g}.
\label{pert}
\end{equation}
First thing to notice is that the linear order in $1/g$  for the energy  
arises from the wave function at $1/g=0$, which is a general 
conclusion as we argue in the next section. Eq.~({\ref{rfunc})
allows us to access also the expansion for the solution to eq.~(\ref{req}). 
However, this expansion does not yield the exact wave function 
to the linear order in $1/g$, since the non-diagonal couplings, $P_{ij},Q_{ij}$,
are proportional to $1/g$ and should be properly taken into account.
However, these couplings can be easily written analytically 
which allows one to determine the corresponding contribution 
numerically from the set of coupled equations~(\ref{phi_r}). 

\section{$N$ Particles}
\label{sec:2}

Here we consider $N=N_A+N_B$ spinless fermions of two types, 
$N_A$ particles of type $A$ and $N_B$ of type $B$, described with the sets 
of coordinates $\{y_i\}$ and $\{x_i\}$, respectively.
Again we assume repulsive zero-range interparticle interaction 
and external confinement, $V(x)$, such that the system 
is described by the Hamiltonian
\begin{equation}
H=\sum_{i=1}^{N_{B}} h(x_i)
+\sum_{i=1}^{N_{A}} h(y_i)
+g\sum_{j,i} \delta(x_i-y_j), \qquad h(x)=-\frac{\hbar^2}{2m}\frac{\partial^2}{\partial x^2} + V(x).
\label{full_ham}
\end{equation}
The Hamiltonian~(\ref{full_ham}) contains delta functions, which 
corresponds to the Schr{\"o}dinger equation 
\begin{equation}
\left[\sum_{i=1}^{N_{B}} h(x_i)
+\sum_{i=1}^{N_{A}} h(y_i)\right]\Psi=E\Psi,
\label{diff_equation}
\end{equation} 
with the following boundary conditions when the particles meet, i.e. at $x_i=y_j$,
\begin{equation}
\left(\frac{\partial \Psi}{\partial x_i}-\frac{\partial \Psi}{\partial y_j}\right)_{x_i-y_j=0^+}-
\left(\frac{\partial \Psi}{\partial x_i}-\frac{\partial \Psi}{\partial y_j}\right)_{x_i-y_j=0^-}=
\frac{2gm}{\hbar^2}\Psi(x_i=y_j).
\end{equation}
We focus on the strongly interacting regime
where $1/g \ll 1$. We write the wave function as
$\Psi = \Psi_0 + \delta \Psi$  with the energy $E=E_0+\delta E$ 
where the normalized wave function $\Psi_0$
is the eigenstate at $1/g=0$ of energy $E_0$.  
Note that to satisfy the boundary condition the wave function $\Psi_0$
should vanish whenever two particles meet, i.e. $\Psi_0(x_i=y_j)=0$.
We can assume that $\langle \Psi_0|\delta \Psi \rangle=0$, where
$\delta \Psi$ solves
the equation 
\begin{equation}
\left[\sum_{i=1}^{N_{B}} h(x_i)
+\sum_{i=1}^{N_{A}} h(y_i) - \delta E -E_0\right]\delta 
\Psi=\delta E \Psi_0,
\label{delta_eq}
\end{equation}
supplemented with the boundary conditions 
\begin{equation}
\left(\frac{\partial \Psi}{\partial x_i}-\frac{\partial \Psi}{\partial y_j}\right)_{x_i-y_j=0^+}-
\left(\frac{\partial \Psi}{\partial x_i}-\frac{\partial \Psi}{\partial y_j}\right)_{x_i-y_j=0^-}=
\frac{2gm}{\hbar^2}\delta \Psi(x_i=y_j).
\end{equation}
Now multiplying eq.~(\ref{delta_eq}) from the left 
with $\Psi_0$ and integrating over the full space except the points with $x_i=y_j$
we get
\begin{align}
& \delta E = \langle \Psi_0|\left(\sum_{i=1}^{N_{A}} h(x_i)
+\sum_{i=1}^{N_{B}} h(y_i)\right)|\delta\Psi \rangle = -
\frac{g\hbar^2}{2m}\sum_{i=1}^{N_A}\sum_{j=1}^{N_B} 
\int\mathrm{d}x_1 ...\mathrm{d}x_{N_B}\mathrm{d}y_1 ... \mathrm{d}y_{N_{A}} \times \nonumber \\
& \left[ \left(\frac{\partial \Psi_0}{\partial x_j}-\frac{\partial \Psi_0}{\partial y_i}\right)_{x_j-y_i=0^+}-
\left(\frac{\partial \Psi_0}{\partial x_j}-\frac{\partial \Psi_0}{\partial y_i}\right)_{x_j-y_i=0^-}\right] \delta\Psi \delta(x_j-y_i) ,
\end{align}
where the second equality is found by integrating 
twice by parts.
The boundary conditions for $\delta\Psi$ demand 
that $\delta E = -K/g + O(1/g^2)$, where
\begin{align}
&K=\frac{\hbar^4}{4m^2}\sum_{i=1}^{N_A}\sum_{j=1}^{N_B} 
\int\mathrm{d}x_1 ...\mathrm{d}x_{N_B}\mathrm{d}y_1 ... \mathrm{d}y_{N_{A}} \times \nonumber \\
& \left[ \left(\frac{\partial \Psi_0}{\partial x_j}-\frac{\partial \Psi_0}{\partial y_i}\right)_{x_j-y_i=0^+}-
\left(\frac{\partial \Psi_0}{\partial x_j}-\frac{\partial \Psi_0}{\partial y_i}\right)_{x_j-y_i=0^-}\right]^2 \delta(x_j-y_i).
\label{slope}
\end{align}
This result was first derived in Ref.~\cite{volosniev}
using the Hellmann-Feynman
theorem. Here we arrive at eq.~(\ref{slope}) directly from the 
non-interacting Schr{\"o}dinger equation with the boundary 
conditions at the points where particles meet. 
Eq.~(\ref{slope}) becomes very useful after
we realize that $\Psi_0$ vanishes at $x_i=y_j$ 
and satisfies the free Schr{\"o}dinger equation otherwise. 
We know that the wave function of $N$ identical spinless fermions $\Psi_A$,
meet these requirements, which means that for each ordering of 
particles, i.e. $x_1<x_2<...<y_{N_A}$, the wave function $\Psi_0$
is proportional to $\Psi_A$.
For identical bosons it was pointed out in 
ref.~\cite{girardeau} that  the ground state wave function at each point is
just an absolute value of  $\Psi_A$. For 
two-component fermions, to obtain $\Psi_0$ 
which is adiabatically connected to the wave function at
large but finite interaction strength, the proportionality 
coefficients for each ordering 
should be chosen to extremize $K$~\cite{volosniev}.
For instance, for three particles in a harmonic trap it was shown~\cite{volfew}
that for the ground state $\Psi_0(x_1<y_1<y_2)=\Psi_A(x_1,y_1,y_2)$,
$\Psi_0(y_1<x_1<y_2)=-2\Psi_A(x_1,y_1,y_2)$, which determines
$\Psi_0$ everywhere, since parity is conserved.  
Now inserting this $\Psi_0$ in eq.~(\ref{slope}) we obtain the same 
energy in linear order in $1/g$ as in~eq.~(\ref{pert}).

\section{Summary and Outlook}
\label{sec:2}

We demonstrate for three harmonically trapped fermions 
an analytical procedure to determine 
the eigenenergy of the system close to the hard core limit 
using the hyperspherical formalism.
It turns out that the leading order correction 
to the energy can be obtained using only the known wave function 
of three spinless fermions. 
In the second part of the paper we present a new detailed derivation of the 
leading order correction 
to the energy for strongly interacting particles~\cite{volosniev}.
This correction depends solely on the wave function of spinless 
fermions, which can be obtained by solving a one-body problem.
Let us now discuss possible extensions 
of this work. 
Here we have only considered leading order corrections 
to the three body problem. However since the $1/g$ expansions 
for the non-adiabatic couplings can be obtained using 
the presented angular wave functions we believe that the higher
order corrections can be obtained numerically in a relatively simple manner. 
The hyperspherical approach discussed here should be of 
great use not only for the presented three-body case but also 
for more particles (bosons and/or fermions), especially in a harmonic trap~\cite{amin}.
Other important extensions are the systems with different masses 
and different interaction strengths
which were successfully treated in homogeneous 
set-ups~\cite{kartavtsev,nirav}.

\end{document}